**Dynamical System Modeling to Simulate Donor T Cell Response to Whole Exome Sequencing-Derived Recipient Peptides: Understanding Randomness in Clinical Outcomes Following Stem Cell Transplantation**

Koparde V,*[1] Abdul Razzaq B,*[2] Suntum T,[2] Sabo R,[3] Scalora A,[2] Serrano M,[1] Jameson-Lee M,[2] Hall C,[2] Kobulnicky D,[2] Sheth N,[1] Sampson J,[1] Roberts C,[2] Buck G,[1] Neale M,[4] Toor A.[2]

* Equally contributing authors.

Center for Biological Complexity, [1] Bone Marrow Transplant Program, Department of Internal Medicine, [2] Department of Biostatistics, [3] Department of Psychiatry and Statistical Genomics, [4] Virginia Commonwealth University, Richmond, Virginia.

Correspondence: Amir A. Toor MD, Professor of Medicine, Bone Marrow Transplant Program, Massey Cancer Center, Virginia Commonwealth University, Richmond, VA 23298. Ph; 804-628-2389. E-mail: amir.toor@vcuhealth.org

**Abstract**

The development of alloreactivity following stem cell transplantation (SCT) in HLA matched donor-recipient pairs (DRP) appears to be random. Minor histocompatibility antigens (mHA) are critical in graft versus host disease (GVHD) pathophysiology however, the quantitative relationship between GVHD and the magnitude of mHA in donor-recipient pairs (DRP) is not known. Whole exome sequencing (WES) was performed on 27 HLA matched related donors (MRD), & 50 unrelated (URD), to identify nonsynonymous single nucleotide polymorphisms (SNPs) with an average 2,490 SNPs were identified in MRD, and 4,287 in URD DRP (p<0.01). The resulting peptide antigens that may be presented on the relevant HLA class I molecules in each DRP were derived *in silico* (NetMHCpan ver2.0) and the tissue expression of proteins these were derived from determined (GTex). MRD DRP had an average 3,626 HLA-alloreactive peptide (AP) complexes with an IC50 of <500 nM and URD, had 5,386 (p<0.01). To simulate a donor cytotoxic T cell response to these AP, the array of HLA-AP complexes in each patient was considered as an *operator* matrix modifying a hypothetical cytotoxic T cell clonal *vector* matrix in which each responding T cell clone's proliferation is quantified by the logistic equation of growth. The simulated steady state T cell clonal response to the AP-HLA complexes in each DRP, accounted for the HLA binding affinity and expression of each AP. The resulting *simulated* organ-specific alloreactive T cell clonal growth revealed marked variability between different DRP. The sum of all T cell clones for common GVHD target organs was: MRD, median 188,821 cytotoxic T cells at steady state (n=26), MUD, 201,176 (n=35), single locus HLA-mismatch MUD: 56,229 (n=10). Despite an estimated, uniform set of constants used in the model for all DRP, and a heterogeneously treated group of patients, overall there was a non-significant trend for higher organ specific T cell counts in patients with GVHD compared to those with none, with weak associations observed for GVHD of the liver. In conclusion, exome wide sequence differences and the variable AP binding affinity to the HLA in each DRP yields a large range of possible alloreactive donor T cell responses possibly explaining the random nature of alloimmune responses observed in clinical practice. Incorporating other quantitative sources of T cell response variability may allow simulations like this to accurately model alloreactivity in SCT in the future.

**Introduction**

Over the last four decades there have been substantial strides made in the optimizing the clinical outcomes following allogeneic stem cell transplantation (SCT). Nevertheless, outcomes such as relapse and graft versus host disease (GVHD) remain difficult to predict in individuals because of the variability observed in the incidence of alloreactivity following HLA matched SCT using conventional GVHD prophylaxis regimens. [1, 2, 3, 4, 5, 6, 7] To an extent this is also true of HLA mismatched transplantation using novel GVHD prophylaxis regimens. [8, 9] Considering that disease responses in allogeneic SCT are often linked to the development of GVHD, it is important to understand the biological basis of the randomness observed in the incidence of alloreactivity. [10] Relapse and GVHD are a function of donor-derived immune reconstitution, which shows many characteristics of a dynamical system, such as logistic growth kinetics and power law distribution of clonal frequencies. [11, 12, 13, 14, 15] This implies that if the quantitative basis of randomness in clinical outcomes can be understood in terms of donor derived immune reconstitution, it will become possible to modify the system to optimize clinical outcomes.

In HLA Matched SCT, recipient antigens, specifically, *minor histocompatibility antigens* (mHA) are presented on the HLA molecules. These mHA released following conditioning or infection induced-tissue injury are taken up by *antigen presenting cells* and presented to non-tolerant donor T cells, which in the event of antigen recognition, proliferate and home back to the target tissue initiating a graft vs. host response. Quantifying the mHA disparity in a SCT donor-recipient-pair (DRP) may therefore allow exploration of the origin of the stochastic nature of alloreactivity. It has been shown that both the magnitude and rate of recovery of donor-derived T cells influence the likelihood of relapse, GVHD and survival. [11] This indicates that T cell responses to recipient antigens may be modeled as dynamical systems, with quantitative rules that govern repertoire evolution under normal circumstances and following transplantation. T cell repertoire is exceedingly complex, it is very likely that the antigenic background will be equally as complex. Therefore the antigenic background in a DRP may be similarly described mathematically as a component of this dynamical system. These antigenic differences in a given transplant DRP may be *partially* determined using whole exome sequencing (WES) of SCT donor and recipient DNA. WES has demonstrated that there is a large library of non-synonymous single nucleotide polymorphisms (nsSNP), and an equally large array of nsSNP-*derived*-peptides which would have different amino acid sequences in each DRP. [16, 17] Of these nsSNPs, those present in the recipient but *absent* in the donor will in theory, yield an array of immunogenic recipient peptides which may trigger donor T cell activation and proliferation, in other words, an array of mHA bound to the 'matched' HLA in the DRP. In aggregate these polymorphisms constitute an *alloreactivity potential* between the specific donors and recipients. In studies done to date, large magnitude of this alloreactivity potential has been observed in all the HLA matched DRP examined. When a dynamical T cell response to this mHA array is calculated marked variation in the simulated T cell response is observed, suggesting that the number and the HLA binding affinity of the mHA may determine that likelihood of alloreactivity. [18] Similar observation reporting association of polymorphisms in the peptide regions of HLA class I molecules support the premise outlined above. [19]

Nevertheless, alloreactivity is a complex clinical state, where patients may have variable manifestations of GVHD impacting different organ systems to varying extent. In this paper the impact of tissue-specific-expression of the proteins that the mHA are derived from in different individuals is examined to measure its variability between unique transplant DRP. This is done using a T cell vector-mHA operator system previously developed [18] with a uniform set of conditions employed to simulate a hypothetical CD8+ T cell response to the *in silico* derived mHA-HLA class I array in 77 HLA matched SCT DRP.

**Methods.**

Patients

Whole exome sequencing (WES) was performed on previously cryopreserved DNA obtained from donors and recipients of allogeneic SCT. Permission for this retrospective study was obtained from the Virginia Commonwealth University's institutional review board, and patient's who underwent transplantation between 2010 and 2014 were retrospectively selected for this analysis. Patients had undergone either 8/8 (n=67) or 7/8 (n=10) HLA-A, B, C and DRB1 matched related (MRD; n=27) or unrelated (MUD; n=50) or haploidentical (n=1) SCT according to institutional standards at Virginia Commonwealth University (VCU) (Supplementary Table 1). HLA matching had been performed using high resolution typing for the unrelated donor SCT recipients; and intermediate resolution typing for class I, and high resolution typing for class II antigens for related donor recipients. A variety of different conditioning and GVHD prophylaxis regimens were used in the patients. HLA class I typing information for the donor and recipient was acquired from the clinical data base.

*Whole Exome Sequencing*

Nextera Rapid Capture Expanded Exome Kit was used to extract exomic regions from the DNA samples, which were then multiplexed and sequenced on an Illumina HiSeq 2500 to achieve an average coverage of ~90X per sample. 2X100 bp sequencing reads were then aligned to the human reference genome using BWA aligner. Duplicate read alignments were detected and removed using Picard tools. Single nucleotide polymorphisms (SNPs) in both the donor and recipients' exomes were determined using GATK HaplotypeCaller walker. GATK best practices were then implemented to filter and recalibrate the SNPs; and store them in variant call file (VCF) format. To identify SNPs unique to the recipient and absent in the donor the results from the GATK pipeline in VCF format were then parsed through the in-house TraCS (Transplant pair Comparison System) set of perl scripts. TraCS traverses through the genotypes of the called SNPs, systematically excluding identical SNPs or editing them to align with the graft-versus-host (GVH) direction thereby generating a new VCF with SNPs for a particular DRP in the GVH direction (SNP present in the recipient, absent in the donor).

The SNPs in this VCF are then annotated either as synonymous or non-synonymous using Annovar. The corresponding amino acid polymorphisms along with flanking regions of each protein are then extracted using Annovar to build peptide libraries of 17-mers for each DRP, with the SNP encoded AA occupying the central position. This library is further expanded by sliding a 9-mer window over each 17-mer such that the polymorphic amino-acid position changes in each 9-mer. The HLA class I binding affinity and IC50 values, which quantify the interactions between all these 9-mers for each DRP and all six HLA class I donor molecules (HLA-A, B and C), NetHMCpan version 2.8 was run iteratively in parallel mode on a linux cluster using custom python scripts. Parsing the NetMHCPan output, unique peptide-HLA combinations present in the recipient but not in the donor, i.e., possessing a GVHD vector, were identified and organized in order of declining mHA-HLA affinity. IC50 (nM) indicated the amount of peptide required to displace 50% of intended or standard peptides specific to a given HLA. Binding affinity is inversely related to IC50, such that, smaller the IC50 value, the stronger is the affinity. The variant alloreactive peptides with a cutoff value of IC50 ≤500 are included in the analyses presented here.

The Genotype-Tissue Expression (GTEx) portal V6 has publicly available expression level information (Reads/kilobase of transcripts/million mapped reads, RPKM values; http://www.gtexportal.org/home/) for a variety of human tissues over a large number of genes. As we know the genes from where the peptides in our DRP peptide library are originating from, we are able to parse in the RPKM values from the GTEx portal for that gene across the whole array of tissues of interest, namely, skin, lung, salivary gland, esophagus, small intestine, stomach, colon and liver. In the DRP where a male recipient had been transplanted from a female donor (n=17), full length available sequences of all proteins encoded by the Y chromosome were curated from NCBI and other resources (see supplementary material). These sequences were then computationally split into 9-mer peptides and their respective binding affinities and IC50 values for the relevant donor HLA antigens were determined in silico using the NetMHCpan software as described earlier. These peptides were then appended to the corresponding DRPs peptide library for all subsequent analyses.

*Computational Methods: Dynamical System Modeling Of T Cell Response To Putative mHA*

It was postulated that the immune effectors and their antigenic targets constitute a *dynamical system,* i.e. an *iterating* physical system, where the variable being studied evolves over time, and the state of any variable at any given time (*t*), is predicated on all the preceding states of the systems. A system demonstrating logistic growth of immune effectors may be described by the general equation

$$\frac{dN}{dt} = rN_t \left(\frac{K - N_t}{K}\right)$$

Here, *dN/dt* is the instantaneous rate of change of the cell population at time *t*, when the population $N_t$ has an intrinsic growth rate, *r* and a maximum size, *K*. This logistic dynamical system constitutes a system of repeated calculations, where the output of the equation for each iteration gives the population $N_t$, as a *function* of time, and becomes the input variable (*argument*) for the next calculation $N_{t+1}$. This system behaves in a *non-linear* fashion, demonstrating sigmoid population growth constrained by feedback. [18, 20, 21]

If SCT is considered as a system of interacting donor T cell clones and recipient antigens, immune reconstitution following SCT may be modeled mathematically. Each T cell clone responding to its target antigen will proliferate, conforming to the *iterating logistic equation* of the general form

$$N_t = \frac{K}{(K - N_{t-1})(e^{-rt}) + 1} \qquad [1]$$

In this equation: $N_t$, is the T cell count at time *t* following transplant (modeled as iterations); $N_{t-1}$ represents the T cell count in the previous iteration; $K$ is the T cell count under steady state conditions reached after several iterations, and represents the maximum T cell count the system would support (carrying capacity); $r$, is the growth rate of the population; and *t* is the number of iterations that the population has gone through (representing time). T cell population, $N_t$, at time, *t* following SCT, is a function of the T cell population $N_{t-1}$, at an earlier time point, *t-1*. This in turn will depend on the T cell population at the outset of transplant $N_0$, and the intrinsic T cell proliferative capacity (growth rate) $r$ governing the growth, such that

$$N_0 \xrightarrow{[r]} N_{t-1} \xrightarrow{[r]} N_t \dashrightarrow^{[r]} K$$

In this model, the final steady state population of all the T cell clones (*n*) ($\sum_1^n K$), and its clonal repertoire may be studied for associations with clinical outcomes.

*Matrices To Model The T Cell Clonal Repertoire: Vector Spaces And Operators*

To apply the logistic equation to describe the evolution of all the individual T cell clones present in an individual at a given time, in other words, to *simulate* the evolution of the T cell clonal repertoire responding to the entire WES derived mHA-HLA library for an individual DRP, matrices are utilized. T cell clones present in an individual may be considered as a set of individual *vectors* in the immune *phase* *space* of that individual. The descriptor, *vector*, in this instance is used to describe the entire set of T cells, with the individual T cell clones representing components of this vector [22]. In this vector, T cell clonality represents direction (since it determines antigen specificity) and T cell clonal frequency, the magnitude of individual vector components. The sum of these vectors will represent the overall T cell vector and its direction, in this simulation, recipient-directed alloreactivity. The T cell vectors may take a number of different configurations in an individual, and the entire range of possible vector configurations will constitute the *immune phase space* for that individual. The T cell vector may be represented mathematically as a single row matrix, with T cell clonal frequencies $TC_1$, $TC_2$, $TC_3$ .... $TC_n$. This T cell clonal matrix may be represented by the term, $\overline{v_{TCD}}$. The frequency of each TC is a natural number, and the direction, its reactivity, represented by the clonality. Phase space will then describe all the potential T cell clonal frequency combinations possible, within a specific antigenic background, with all the potential directions of the T cell reactivity. In the case reported here, the vectors are all in the direction of graft vs. host alloreactivity, since they proliferate after encountering mHA-HLA. In contrast, the infused T cell clonal repertoire will represent a steady state T cell clonal frequency distribution in the normal donor, presumably made up of mostly self-tolerant, pathogen specific T cells, and without any auto-reactive T cells. For the computations reported here, each T cell clone has a $N_0$ value of 1, and is alloreactive.

The T cells infused with the transplant encounter a new set of antigens (both recipient and pathogen). These antigens presented by the recipient (or donor-derived) antigen presenting cells constitute an *operator*, a matrix of targets in response to which the donor T cells may proliferate [23]. This operator changes the magnitude and direction (clonal dominance) of the infused T cell repertoire, as individual T cell clones in the donor product grow or shrink in the new HLA-antigen milieu, transforming the T cell vector. The putative mHA making up the alloreactivity potential constitute a matrix, which may be termed an *alloreactivity potential operator,* $\mathbb{M}_{APO}$. This operator will consist of the binding affinity of the unique variant peptide-HLA complexes likely to be encountered by donor T cells in an individual recipient. Following SCT, the donor T cell clonal frequency shifts according to the specificity of the TCR and the abundance and reactivity of the corresponding antigen. This corresponds to the operator modifying the <u>original</u> *donor* T cell vector, $\overline{v_{TCD}}$ as the system goes through successive iterations to a <u>new</u> *recipient* T cell vector, $\overline{v_{TCR}}$, according to the following relationship

$$\overline{v_{TCD}} \cdot \mathbb{M}_{APO} = \overline{v_{TCR}} \qquad [2]$$

*Applying the Logistic Growth Equations to Vector-Operator Systems*

A central assumption in this model is that the steady state TC clonal frequency ($K$) of specific TCR bearing clones, and their growth rates ($r$), are proportional to the binding affinity of their target mHA-HLA complexes. This is *approximately* represented by the reciprocal of the *IC50* (test peptide concentration in *nM* required to displace a standard peptide from the HLA molecule in question) for that specific complex. This is because the strength of the binding affinity will increase the likelihood of T cell-APC interactions occurring, thus influencing the driving the rate of this interaction. While each TCR may recognize another mHA-HLA complex with equal or lesser affinity, for the sake of simplicity, an assumption of non-recognition of other mHA-HLA complexes was made for this model. This makes it possible to specify an *identity matrix*, Matrix A, with unique mHA-HLA complexes for a SCT-DRP and the corresponding TCR, where $TCR_x$ recognizes $mHA_x$-HLA (1) and not others (0). In addition to the assumption that there is a first signal of TCR recognition, a second signal immune responsiveness is assumed in this matrix.

Matrix I. Effect of alloreactivity operator $\mathbb{M}_{APO}$ on T cell vector $\overline{v_{TCD}}$. Successive iterations ($t$) modify the vector to $\overline{v_{TCR}}$. In this simplified model, $mHA_x$-HLA only effects $TC_x$ and so on. Matrix below represents a single iteration $t$.

| SCT | | $\mathbb{M}_{APO}$ | | | |
|---|---|---|---|---|---|
| | | $mHA_1$ HLA | $mHA_2$ HLA | $mHA_3$ HLA | $mHA_n$ HLA |
| $\overline{v_{TCD}}$ | $TC_1$ | 1 | 0 | 0 | 0 |
| $\downarrow t$ | $TC_2$ | 0 | 1 | 0 | 0 |
| $\overline{v_{TCR}}$ | $TC_3$ | 0 | 0 | 1 | 0 |
| | $TC_n$ | 0 | 0 | 0 | 1 |

The *alloreactivity matrix* modifies the donor T cell clonal vector infused with an allotransplant, mapping it to the recipient T cell vector, as the T cell clones with unique TCR encounter the corresponding mHA-HLA complexes they proliferate conforming to the logistic equation [1]. In the logistic equation $K$ for each T cell clone will be proportional to the approximate binding affinity ($1/IC50$) of the corresponding mHA-HLA complex ($K^{1/IC50}$). The parameter $r$ is a function of the binding affinity (by increasing the TCR-mHA-HLA interaction time) and the intrinsic T cell proliferation capacity. Equation 1 must therefore be modified for unique TC clone $TCx$, responding to $mHAx$-$HLA$, as follows:

$$N_{t\ TCx} = \frac{K^{1/IC50mHAx}}{(K^{1/IC50\ mHAx} - N_{t-1\ TCx})(e^{-rt(1/IC50mHAx)}) + 1} \quad [3]$$

This general equation gives instantaneous T cell counts in response to antigens presented, regardless of tissue distribution. In the alloreactivity, vector-operator identity matrix, the values 1 or 0 in each cell are multiplied by the product of equation 3 for each T cell clone. This is depicted in Matrix II.

Matrix II. Matrix illustrating a single iteration of the alloreactivity operator $\mathbb{M}_{APO}$ on T cell vector $\overline{v_{TCD}}$. Each cell in the matrix calculates the value of $TC_x$ in response to $mHA_x$-HLA, final repertoire is determined by solving the matrix.

| SCT | | $\mathbb{M}_{APO}$ | | | |
|---|---|---|---|---|---|
| | | $mHA_1$ HLA | $mHA_2$ HLA | $mHA_3$ HLA | $mHA_x$ HLA |
| $\overline{v_{TCD}}$ ↓ $t$ | $TC_1$ | $N_{t\ TC1} = \frac{K^{1/IC50mHA1}}{(K^{1/IC50\ mHA1} - N_{t-1\ TC1})(e^{-rt(1/IC50mHA1)}) + 1}$ | 0 | 0 | 0 |
| $\overline{v_{TCR}}$ | $TC_2$ | 0 | $N_t TC_2$ | 0 | 0 |
| | $TC_3$ | 0 | 0 | $N_t TC_3$ | 0 |
| | $TC_x$ | 0 | 0 | 0 | $N_t TC_x$ |

The T cell response to each mHA-HLA complex is determined over time, $t$, by iterating the system of matrix-equations. In this *alloreactivity matrix*, the IC50 of all the alloreactive peptides with a GVH direction (present in recipient, but absent in donor) constitutes the operator; the sum of $n$ T cell clones $\sum_1^n TC$, at each time point will represent the magnitude of the vector $\overline{v_{TCR}}$ at that time $t$. In this system, when considering the effect on infused donor T cell vector, depending on antigen affinity, T cell clones present in abundance may be down-regulated if antigen is not encountered. On the other hand, clones present at a low frequency may expand upon encountering antigen, *transforming* the vector over time from the original infused T cell vector. In summary, the alloreactivity operator determines the change in donor T cell vectors, following transplantation, in an iterative fashion *transforming* it to a new configuration, over time $t$, based on the mHA-HLA complexes encountered in the recipient and their affinity distribution (*and antigen abundance, vide infra*). Therefore post-SCT immune reconstitution may be considered as a process in which T cell clonal frequency vectors are iteratively multiplied by the minor histocompatibility antigen matrix operator and this results in transformation of the vector over time to either a GVHD-prone alloreactive or to a tolerant, pathogen-directed vector. This may be visualized as thousands of T cell clones interacting with antigen presenting cells, an example of an interacting dynamical system. [11]

*Competition between T Cell Clonal Populations*

Each of the T cell clones is behaving like a unique population, therefore *competition* with other T cell clones in the set of all T cell clones must be accounted for in the logistic equation to determine the magnitude of the unique clonal frequencies as the model iterates simulating T cell clonal growth over time. This may be done using the *Lotka-Volterra* model for *competing populations*, which accounts for the impact of population growth of multiple coexisting populations [18, 24, 25]. This is accomplished by modifying the expression $N_{t-1}$ in equation 3 for each clone, by taking the sum of $N_{t-1}$ for all the other competing T cell populations when calculating $N_t$ for each clone. Each clone's $N_{t-1}$ is weighted by a correction factor $\alpha$ for its interaction with the T cell clone being examined. Given the central role for the target mHA-HLA complex's $IC50$ in determining the T cell frequency for each clone, $\alpha$ for each clone is calculated by dividing the $IC50$ of the competing T cell clone with the test clone (note the use of $IC50$ instead of $1/IC50$). This implies that T cell clones recognizing mHA-HLA complexes with a higher binding affinity will have a disproportionately higher impact on the growth of T cell clones binding less avid mHA-HLA complexes and vice versa. The resulting square matrix, Matrix III, will have 1 on the diagonal, and values <1 above the diagonal and >1 below it.

Matrix III. Matrix illustrating the relative effect of antigen binding affinity on the T cell clonal interaction between different clones. Successive cells in each row of the matrix calculate the effect of $TC_i$ on T cell being studied, $TC_x$. This generates a weighting factor, $\alpha$, which modulates the impact of population of $TC_i$ on the growth of $TC_x$.

|        | $TC_1$ | $TC_2$ | $TC_3$ | $TC_n$ |
|--------|--------|--------|--------|--------|
| $TC_1$ | $IC50_1/IC50_1$ | $IC50_1/IC50_2$ | $IC50_1/IC50_3$ | $IC50_1/IC50_n$ |
| $TC_2$ | $IC50_2/IC50_1$ | $IC50_2/IC50_2$ | $IC50_2/IC50_3$ | $IC50_2/IC50_n$ |
| $TC_3$ | $IC50_3/IC50_1$ | $IC50_2/IC50_3$ | $IC50_3/IC50_3$ | $IC50_3/IC50_n$ |
| $TC_n$ | $IC50_n/IC50_1$ | $IC50_n/IC50_3$ | $IC50_n/IC50_3$ | $IC50_n/IC50_n$ |

To account for *n* competing T cell populations, equation 3 will be modified as follows

$$N_{t\ TCx} = \frac{K^{1/IC50mHAx}}{(K^{1/IC50\ mHAx} - \sum_1^n N_{t-1\ TCi} \cdot \alpha_i)(e^{-rt(1/IC50mHAx)}) + 1} \qquad [3.1]$$

Where for the T cell clone $x$, $N_t$ depends on the sum of the $n$ T cell clonal frequencies at the previous iteration, with the α of each T cell clone $i$, with respect to the T cell clone $x$, modifying the effect of the frequency of the $i$th clone on $TCx$. By contrast, the parameter $r$ can have either positive or negative values depending on whether the ambient cytokine milieu, either stimulates (-$r$) or suppresses (+$r$) growth.

The direction of the T cell clones will be determined by antigen specificity, i.e. whether their TCR recognize the recipient's mHA-HLA (are alloreactive) or not (are tolerant), such as pathogen directed T cell clones. For example, a tolerant, non-autoreactive set of T cell clones (donor vector, $\overline{V_{TCD}}$) may be transformed to a predominantly alloreactive set of T cell clones ($\overline{V_{TCR}}$), after interacting with the mHA-HLA complex, alloreactivity operator encountered in the recipient ($\mathbb{M}_{APO}$).

*Accounting for tissue expression of proteins*

The peptides discussed in the afore mentioned derivation are generated from proteins expressed in target tissues, as such the level of protein expression will determine the magnitude of the peptide specific T cell response. The higher the protein expression, the more peptide molecules and the greater the HLA presentation with an ability to stimulate a larger T cell response (clonal frequency). Therefore the parameter $K$ in the above equations may be modeled as a multiple of the level of protein expression. The tissue expression of proteins is estimated by RNA sequencing techniques, and is measured in either transcripts per million (TPM) or reads per kilobase of transcript per million mapped reads (RPKM). The values available from the public data base GTEx, range from 0 to >$10^4$ and constitute a coefficient in the definition of K. This Modifies Equation 3.1,

$$N_{t\ TCx} = \frac{(RPKM_x.K)^{1/IC50mHAx}}{\left((RPKM_x.K)^{1/IC50\ mHAx} - \sum_1^n N_{t-1\ TCi}.\alpha_i\right)(e^{-rt(1/IC50mHAx)}) + 1} \quad \ldots\ [3.2]$$

In equation 3.2, the tissue expression of the protein from which the target peptide, $mHAx$ is derived is incorporated as a K multiplier when calculating a tissue-specific T cell response. Thus tissue-specific alloreactivity potentials may be simulated for each of the relevant GVHD target tissues, by substituting equation 3.2 into Matrix II.

In applying this model to exome sequence derived, alloreactive-peptide-HLA binding patient data an IC50 cutoff value of ≤500 nM, and an RPKM value of ≥1 were chosen to study the differences between patients. The T cell repertoire simulations were then run in MATLAB (Mathworks Inc., Natick, MA), utilizing the above model (See Supplementary Mathematical Methods and Program).

The sum of all T cell clones $\sum_1^n TC$ for each specific organ of interest and the grand total of the organs studied (termed sum of all clones) were used to represent the T cell vector magnitude of the alloreactive T cells at steady state for each patient.

*Statistical Methods*

Estimated T cell counts are summarized per organ and in total with means and standard deviations, both overall and by GVHD classification. In addition, variability measures of estimated T cell counts across organs within individuals (standard deviation, minimum, maximum, and range) are summarized in the same manner. The estimated T cell counts and variability measures are compared between GVHD and Donor Type groups using the Wilcoxon rank sum test. Cox proportional hazards models are used to examine associations between both the estimated T cell counts and variability measures with patient outcomes (GVHD, survival, relapse and relapse-free survival). The MEANS, NPAR1WAY and PHREG procedures in the SAS statistical software program (version 9.4, Cary, NC, USA) are used for summaries and analyses. All estimated T cell counts are divided by 1000 to simplify the presentation of results.

**Results.**

*Exome sequencing*

Cryopreserved DNA samples from 78 donors and recipients of allogeneic SCT were sequenced, demographic details of the patients are given in supplementary table 1. Whole exome sequencing revealed a significantly larger number of SNPs, both synonymous as well as non-synonymous in recipients of MUD SCT when compared with MRD recipients (Table 1). Further, a majority of these SNPs were nonconservative, i.e. likely to substitute an AA with different physico-chemical properties, so liable to bring about a conformational change. In patients who either did or did not develop GVHD, an average of 2456 and 2474 nsSNPs were identified in recipients of MRD, while the corresponding numbers for MUD recipients were 4536 and 3845 nsSNPs. In the one haploidentical transplant recipient, there were 3231 ns SNPs. Notably, the transition/transversion ratio in the WES data was as expected between 2.1 and 2.5 for the entire data set (Supplementary Fig 1).

*In silico derivation of mHA-HLA complexes from exome variation*

The $nsSNP_{GVH}$ data arrays then served as the basis to computationally determine peptide sequences. For each variant amino acid (AA) resulting from a $nsSNP_{GVH}$, the flanking 8 AA on either side were determined and the resulting nonamers, with the variant AA occupying all nine positions derived for each variant AA. The binding affinity of each of these peptides was determined to the HLA class I (A, B and C) molecules in each pair. This yielded a large array of unique peptide-HLA complexes, in other words *putative* or *potential* minor histocompatibility antigens (mHA) in each individual; once again a significantly larger number of peptides of both presented and strongly bound, were observed in MUD recipients when compared with MRD (Table 1). A similar number of peptides with an IC50 <500nM, 3826 and 3406 pmHA, were recorded in MRD patients either with or without GVHD respectively; corresponding numbers were 895 and 778 mHA with an IC50 <50nM for those patient groups. For the MUD SCT pairs (n=50), the corresponding numbers were 5672 and 4876 pmHA with an IC50 <500 nM and 1210 and 1071 with an IC50 <50 nM in patients with or without GVHD. Notably these numbers do not include the Y chromosome derived peptides presented by the class 1 HLA molecules in male recipients of female donors. The haploidentical donor had 2448 peptides with an IC50 of <500 nM out of a total 57957 peptides; of these 484 were strongly bound, when using the recipient HLA for performing the calculation.

*Simulating recipient tissue specific donor T cell responses*

Following determination of the mHA with various binding affinity distributions, the T cell responses to these antigens were simulated using the T cell vector- alloreactivity operator model [Equations 2 & 3.2]. The underlying conditions used in these calculations were as follows: for each mHA there was a single T cell present at the first iteration, i.e. $N_t$ = 1 at $t$ = 1. The $K$ in equation 3.2 was set at 1000,000 cells (note: this value substitutes for organ mass, lympho-vascular supply etc. without accounting for differences between organs), and $r$ was set at 1.5 for all the simulations (note: in vivo this may vary with state of inflammation). All the mHA were included in the simulations with competition, and included Y chromosome encoded peptides which bound respective HLA in the male recipients of female donors. The program accounted for both IC50 and tissue expression measured as RPKM for each mHA to simulate donor T cell responses to organs of interest, with skin, salivary gland, esophagus, stomach, small intestine, colon, liver, spleen, blood, blood vessel, lung included in the analysis. Individual clonal growth was variable depending on the HLA binding affinity of the peptide, as well as the tissue expression of protein, however the relationship was not linear (Figure 2). T cell clones responding to low expression/low binding affinity were suppressed and had negative values and recovered over time (iterations of the system). The T cell clonal growth for each organ's alloreactive peptides was summed to give the simulated alloreactive T cell count for each organ in each individual. The average value for each organ at steady state (average of the iteration # 401-500) was determined and is shown for each patient (Figure 3A) and depicts the magnitude of organ-specific alloreactive T cell response that may be seen in different DRP, accounting for their HLA types and exome sequence variation. The simulated organ-specific T cell counts responding to the alloreactivity operator demonstrate marked variability in both HLA matched related and unrelated donors, both between different DRP (Figure 3B) as well as within DRP (Figure 3C). This is true for individual organ simulations, as well as the sum of these simulations (Figure 3D), which reflects the variation in the organ specific T cell responses.

When the sum of all organ specific T cell clones (Σ TC) was examined in different donor types, no significant differences were observed in any of the categories examined; MRD: median 188,821 ΣTC at steady state (n=26) vs. MUD: 201,176 (n=35) vs. HLA mm donor: 56,229 (n=10). However, while the differences were not statistically significant between these retrospectively chosen groups, there was a trend supporting quantitative differences between different donor types. Estimated T cell counts are summarized by donor type in Supplementary Table 2, where for every organ (and overall) the mean expected T cell counts and estimated variabilities are larger in patients with unrelated donors than they are in patients with matched related donors, however none of these differences were significant (all p-values > 0.3), most likely due to the high variability of these estimates. No significant differences were observed when other biological features such as gender mismatch in the DRP (F to M: 270,613 (n=17), M to F: 135,064 (n=14) vs. gender match (F to F: 180,574 (n=16), M to M: 114,932 (n=26), and difference in races, (AA to AA: 113,245 (n=13), vs. C to C: 215,337 (n=47), vs. race mm: 57,938 (n=6)) were studied. This corroborates well with the relatively weak effect of these biological differences in the development of GVHD in the setting of HLA matching. [26, 27] However the small sample size and heterogeneity of our group in terms of conditioning and GVHD prophylaxis may also contribute to the variability observed in clinical outcomes.

*Clinical association of tissue-specific alloreactivity potential*

The simulated T cell counts are summarized in Supplementary Table 3, where for every organ (and overall) the mean simulated T cell counts were larger in patients who eventually developed any form of GVHD, than they were in patients who did not develop GVHD, though none of these differences were significant. As can be seen in Supplementary Table 4, that while none of the organ-specific simulated T cell counts were associated with GVHD (all p-values between 0.05 and 0.15), the variability (standard deviation) among organ-specific T cell estimates in individuals is significantly associated (HR = 1.08, 95% CI: 1.01, 1.15, p = 0.0275) with GVHD, as was the maximum organ-specific T cell estimate (HR = 1.01, 95% CI: 1.00, 1.02, p-value = 0.0462) and the range between the maximum and minimum organ-specific t-cell counts (HR = 1.02, 95% CI: 1.00, 1.05, p-value = 0.0276). These results imply that increased variability of T cell counts among organs are associated with a minimal increase in the risk of GVHD. Similar analyses for survival, relapse and relapse-free survival were not significant for any of the organ-specific T cell counts or the variability measures (all p-values > 0.5). While these are weak associations at best, it is noteworthy that despite an estimated, uniform set of constants used in the model for all DRP, heterogeneously treated group of patients, overall there was a trend for higher organ specific T cell counts in patients with GVHD compared to those with none.

Cox proportional hazards models for acute and chronic GVHD are presented in Supplementary Table 5, and demonstrated no significant associations between expected t-cell counts and variability measures with acute GVHD (all p-values > 0.05). For chronic GVHD, only the expected T cell counts from the liver were significantly associated, with a slight positive relationship (HR = 1.01, 95% CI: 1.00, 1.03, p-value = 0.0411). The standard deviation among organ-specific t-cell estimates is significantly (p = 0.0404) and positively associated (HR = 1.08, 95% CI: 1.01, 1.16) with GVHD, as was the range between the maximum and minimum organ-specific T cell counts (HR = 1.03, 95% CI: 1.01, 1.05, p-value = 0.0266). These results imply that increased variability of t-cell counts among organs are associated with slight increases in the risk of chronic GVHD.

There were no associations between organ-specific expected T cell counts and the organ-specific occurrence of acute GVHD (all p-values > 0.5). Only expected T cell counts in salivary glands was significantly and positively associated with oral chronic GVHD (HR = 1.02, 95% CI: 1.01, 1.04, p = 0.0131); all other organ-specific expected t-cell counts were not associated with organ-specific chronic GVHD (p-values > 0.09).

**Discussion.**

Alloreactivity following SCT is a complex disorder dependent on a number of factors such as degree of HLA matching, [28, 29, 30] intensity of immunosuppression [31, 32, 33] and the graft source/T cell composition of the graft. [34, 35] However, even within uniformly HLA matched and immunosuppressed patients with similar disease biology, outcomes remain variable and subject to laws of probability. In this paper the magnitude of difference in putative alloantigen presentation between SCT DRP is explored, accounting for the HLA make up and protein coding differences between individual donors and recipient. Marked variation in the likelihood of alloantigen presentation is observed in the simple mathematical model utilized. Further despite the limited nature of the model used, a trend supporting association of the alloreactive simulations with clinical GVHD is observed.

Logically, the binding affinity and tissue expression of the antigens are critical parameters when assessing the T cell response to these antigens. In this paper findings of an earlier model are extended to simulate T cell responses to the entire array of alloantigens that the donor T cells may encounter in a recipient to determine tissue specific alloreactivity potential. Similar to the previous study, the findings reported herein demonstrate a very large degree of variability between the simulated T cell responses to antigens expressed in specific organs between unique pairs. Furthermore, while, the association between clinical GVHD and the simulated T cell responses is not established, there are intriguing associations observed which represent a new area of investigation which might yield interesting results in the future.

The inability to identify strong clinical association between the magnitude of the simulated T cell response to mHA array in a DRP and GVHD, may be explained by, (a) uniform simulation conditions applied in the model, despite a heterogeneously treated patient cohort, with limited numbers in each group, (b) only 9-mer oligo peptides bound to HLA class I considered in the simulations, and those too without any post translational modification (phosphorylation, glycosylation) and protein cleavage site information incorporated. Beyond these obvious considerations there are other important considerations that needs to be factored in. Logic would dictate that when proteins are cleaved, there will be a host of peptides generated for antigen presentation, both alloreactive, such as the ones reported here, as well as non-alloreactive peptides which will also bind HLA with varying levels of affinity, as the alloreactive peptides do. Furthermore these peptides will likely be far more numerous than the alloreactive ones, and may constitute a significant competitive barrier to the presentation of alloreactive peptides. This may be modeled numerically using the notion of *combinatorial probability*. [36, 37] In other words, if there are $n$ total peptides which might be presented on $k$ HLA molecule there are,

$$\binom{n+k-1}{k} = \frac{n+k-1!}{k!(n-1)!}$$

possible combinations of peptides which might be presented by these HLA molecules, including duplication of peptides and without regard to order of peptides presented. For example, if 10 peptides are to be presented by 4 HLA molecules the total number of possible peptide-HLA combinations will be, $\binom{13}{4} = 715$. Now, of these, if 3 peptides are alloreactive (mHA), the probability of a favorable outcome (NO GVHD) will be enhanced if the other 7 non-alloreactive peptides are presented, the number of possible combinations allowing that will be $\binom{10}{4} = 210$. Therefore the probability of having a peptide combination presented by one of the four HLA molecules, which will not include an alloreactive peptide in this mix of 3/10 alloreactive peptides, will be approximately

$$\frac{\binom{10}{4}}{\binom{13}{4}} = 0.293$$

This means that the remaining ~70% of the peptide-HLA combinations will contain one or more alloreactive peptides, if this combination of alloreactive and non-alloreactive peptide combinations were to be presented. Depending on the binding affinity of the peptides, and the potential of the alloreactive peptide to stimulate a T cell response, and the presence of a T cell clone recognizing that specific peptide HLA combination, alloreactivity may or may not be triggered. In real life however the number of non-alloreactive peptides is likely high resulting in a smaller ratio of alloreactive peptides being presented. Expression levels of HLA molecules ($k$ *in the above equations*) have been shown to impact alloreactivity in, supporting this quantitative reasoning. [38, 39] Nevertheless this quantitative consideration adds an important variable to determining the likelihood of alloreactivity developing; i.e. the variability in the HLA binding affinity of the many non-alloreactive peptides (not accounted for in the analysis presented in this paper). Further variability in this system is introduced by the relative abundance of each protein from which the peptides are derived. This too will modify the likelihood of peptide binding to HLA, due to competition. This discussion demonstrates that while dynamical systems modeling reproduces T cell patterns seen in patients, randomness of the exome mutations, and subsequent peptide-HLA interactions makes the development of alloreactivity a *stochastic*, dynamical process. This means that while patterns of reactive immune responses may be mathematically described, each of these patterns is associated with a certain probability of occurring and thus bringing about the associated clinical outcome. In other words, HLA matched patients with a higher simulated T cell count will likely be at greater risk of GVHD under certain uniform conditions of immunosuppression and vice versa.

Another source of difference between an idealized simulation and the 'real world' will be the interaction between APC and T cells. Both the cell populations proliferate in response to inflammatory stimuli, with the APC proliferation driving T cell growth. This interaction too may be modeled, using the matrix based vector operator model. The results reported above are based on the notion that the APC matrix operator ($\mathbf{M}_{APO}$) has a constant effect of the donor T cell vector, in other words antigen presentation is a constant function of time, however in life this is not the case. APC themselves respond to inflammatory signals and as previously demonstrated follow logistic growth and decline after stem cell transplantation, consequently antigen presentation follows a crescendo-decrescendo course over time (particularly with infections). [40, 41] This model may be built into this system of vector operator immune response modeling. In this instance consider the scenario; tissue injury following conditioning therapy ensues, following which there is uptake of recipient antigens by the transplanted APC (presumably of donor origin) proliferating in response to the cytokines released as a consequence of the insult. The proliferating APC take up antigen and present the mHA (and non-mHA) and migrate to the regional lymph nodes. As these monocyte populations arrive in the regional nodes, the T cells corresponding to the mHA start proliferating and migrate out of the lymph node and homing to the target tissue, where they now induce further inflammatory injury, further expanding the APC proliferation, until the APC population reaches a steady state. Once the tissue damage is complete (or infection resolved), the APC will decline and a steady state memory T cell complement left behind. This is a positive feed back loop which 'self regulates' as steady state values are reached for the APC population, with an eventually decline in the antigen presentation (as pathogen numbers decline, or maximal tissue injury runs its course). Similar to T cell clonal proliferation, APC proliferation may be described by equation 1. The progressive rise and fall in the number of APC, (and consequent waxing and waning antigen presentation over time) may then be described as a vector, and its impact in T cell population be recorded by taking the dot product of this and the T cell vector, in other words multiplying each iteration of the T cell vector as it transforms. The logistic variable describing the impact of APC growth and decline on T cell clonal proliferation is given by the expression,

$$v_{APC} = \left(1 - \left(\frac{N_{t\,APC}}{K_{APC}}\right)t\right) + 1$$

This variable is at its highest value, 2, in the beginning of the reaction ($N_{t\,APC}/K_{APC}$ approaches 0) but as the reaction proceeds over time $t$, it approaches 1, as $N_{t\,APC}$ approaches $K_{APC}$. Here 1 represents the binary (0/1) condition of T cell clone specificity for target mHA in $M_{APO}$, in real world terms the steady state capability of the T cell to proliferate when encountering the relevant antigen. Assuming that an independent $v_{APC}$ exists for each antigen in the $M_{APO}$, the effect of APC growth and subsequent mHA presentation may be obtained by the dot product of the two vectors, $v_{APC}$ and $v_{TCR}$. In other words, multiplying each iteration of the T cell vector with each iteration of this APC vector (the operation of multiplying 2 column matrices), transforms the donor T cell vector,

$$v_{TCR}' = v_{TCR} \bullet v_{APC} \bullet cos\theta$$

This equation implies that in the beginning of an inflammatory response there is a positive feedback response from proliferating APC, which amplifies the T cell response $v_{TCR}'$; eventually both decay to a steady state level. The angle $\theta$ between the two vectors is 0 ° ($cos\ 0 = 1$) because they have the same direction, in other words the TCR recognize the mHA-HLA complex being presented by the APC. This effect of $v_{APC}$ on $v_{TCR}$ can be applied to the entire T cell repertoire and is shown in Figure 4A for a single T cell clone and may be generalized to the whole repertoire (Figure 4B). As can be seen these graphs recapitulate the T cell response amplification commonly observed in response to antigen stimulation, [40, 41, 42] and can be depicted in the model illustrated in Figure 4C. In practice, this interaction depending on the presence or absence of inflammation in the tissues being studied, will significantly modify the alloreactivity operator and lead to variability in T cell proliferation observed. The weak associations of the T cell simulations with the clinical findings reported in this paper, may therefore be explained by absence of information regarding the overall inflammatory state in each DRP. This can certainly be modeled in the future iterations of such work.

Another source of clinical variability that is not accounted for in this model is the make up of the graft both in terms of effector, as well as regulatory T cells (Treg). While the notion of absence of antigen directed clonal T cells impacting the predictive power of any such model is straight forward, the effect of regulatory immune cell populations such as Treg requires some calculation to understand. Treg recognize self-antigens and secrete anti-inflammatory cytokines (IL-4, IL-10) in response, diminishing the proliferation of the antigen directed CD8+ effector T cells. [43] This may be easily accounted for in this model by considering that the growth parameter $r$ (given the uniform value of -1.5 in this analysis) represents the proliferative effect of the cytokine milieu. In this instance the more negative its value (induced by pro-inflammatory cytokines), the larger the growth promoting effect, and the less negative or positive the value (anti-inflammatory cytokine effect) the more growth will be impeded or stopped, as can be seen in Figure 5, where a change in the magnitude of r produces dramatic decline in the T cell growth curves. Once again these effects are not accounted for in the model presented here but may be easily included. Nevertheless, the robustness of the vector operator model of quantifying T cell response and its eventual clinical utility are indicated by its robustness in quantifying a variety of immune phenomenon noted above.

Current clinical practice is to identify the donors is based on the degree of major histocompatibility locus matching. Based on this, immunosuppressive regimens are chosen; unrelated donor transplant recipients frequently receive ATG in addition to the standard calcinuerin inhibitor + methotrexate, haploidentical transplant recipients receive post transplant cyclophosphamide and so on. In the absence of early GVHD onset, immunosuppression intensity is gradually reduced and eventually withdrawn over a predetermined period of time, usually spanning four to six months, depending on the relapse risk of the underlying malignancy. Dynamical system modeling of immune reconstitution, based on whole exome sequencing data from donors and recipients may enable the development of patient specific immunosuppressive regimens post transplant, by more precisely calibrating the GVHD risk that unique donors might pose to that patient. Given the advances in *next generation sequencing*, as well as computing technology it is conceivable that in time patients and their prospective donors will have their HLA type and alloreactivity potential determined by whole exome sequencing. A donor with optimal alloreactivity potential will be identified and a GVHD prophylaxis regimen of optimal intensity utilized to achieve maximal *likelihood* of good clinical outcome. In so doing, dynamical systems understanding of alloimmune T cell responses will *attenuate* the unpredictability that the current largely probability based models of outcomes prediction expose patients to. The dynamical systems analysis of antigenic variation also explains the randomness at hand in human immune response to disease, either infectious or neoplastic. This understanding has the potential to impact areas of investigation beyond transplant alloreactivity, potentially influencing cancer immunotherapy, autoimmune disease and infectious disease. Randomness within the dynamical system will remain a problem for the foreseeable future, but a model such as this is a step towards gaining a quantitative understanding of complex immune responses.

In conclusion this model partially explains why immune responses are seemingly random and difficult to accurately predict, akin to the quantum uncertainty principle, you can accurately measure either a particle's position or its velocity, never both.  Similarly, given the complexity at hand in immune responses, we will not be able to precisely quantify the likelihood of alloreactivity, but with a quantitative understanding give a more accurate estimate of the probability distribution.

**Figures.**

**Figure 1.** TRaCS, computational algorithm to determine tissue specific alloreactive peptide mHA.

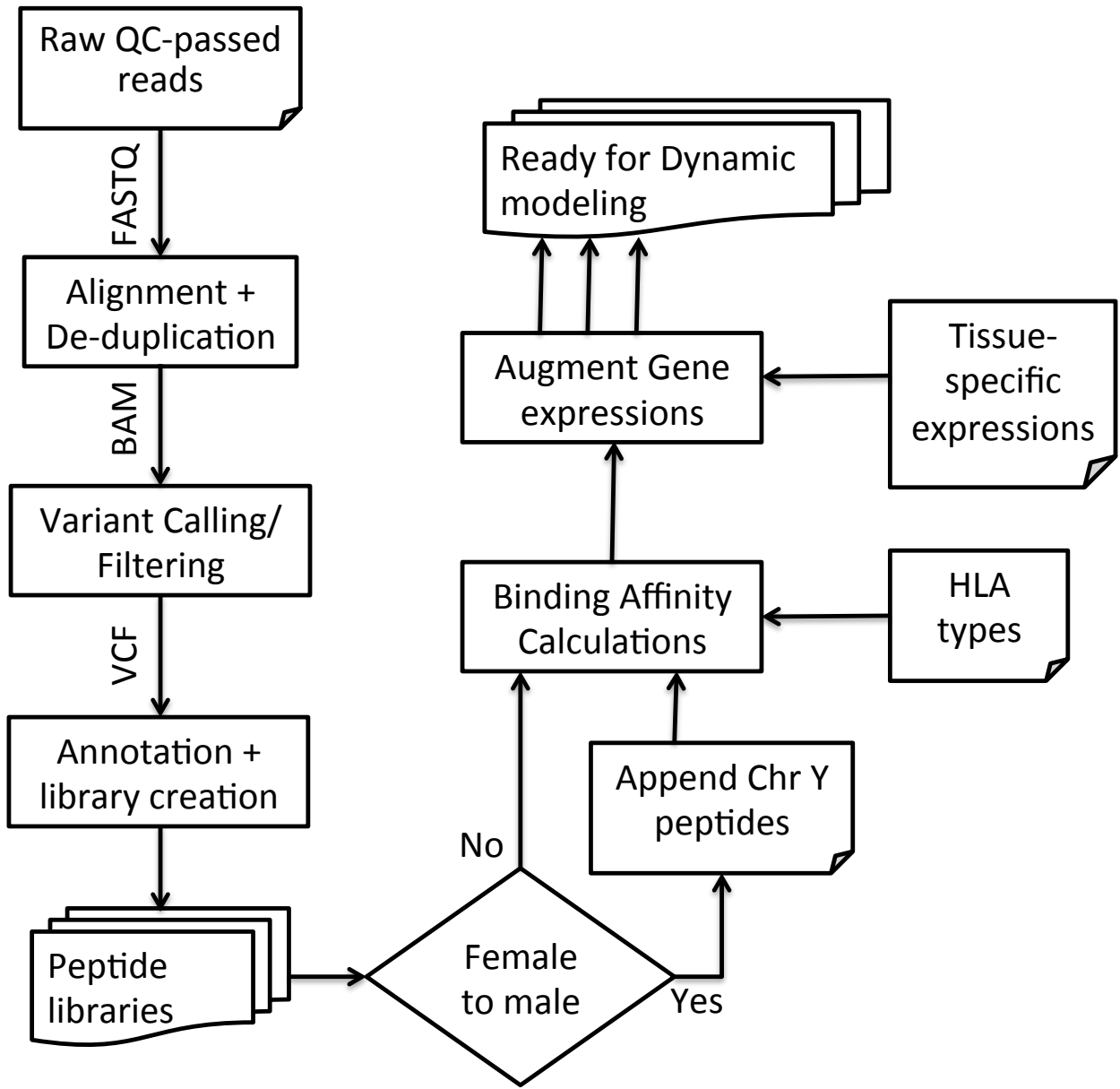

**Figure 2.** Individual T cell clonal growth simulations accounting for peptide-HLA complex binding affinity and protein of origin RPKM. Increased T cell frequency (Y-axis) seen if the protein is expressed at a higher level. IC50/RPKM given for each T cell clone.

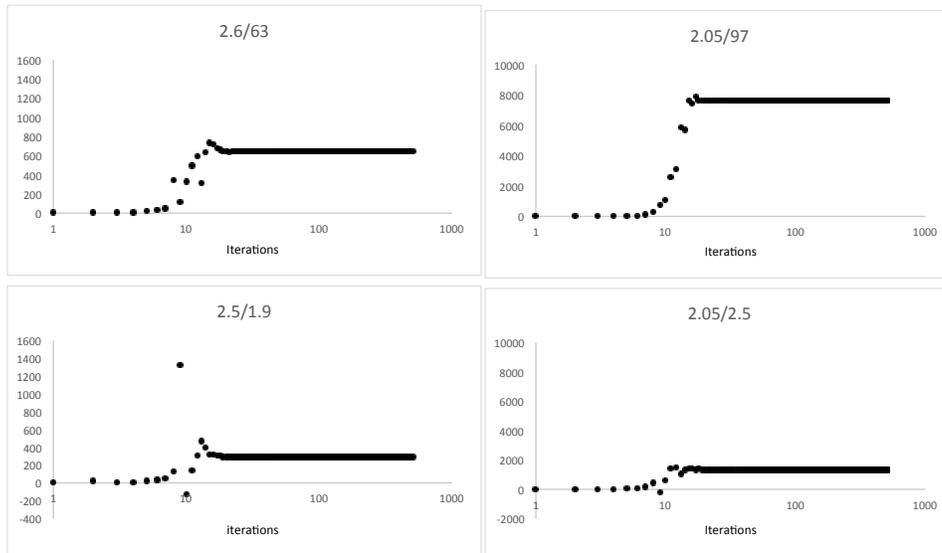

**Figure 3A.** Tissue specific simulated alloreactive T cell counts in recipients of MRD (Black line) and unrelated donors (Blue line). These values represent the sum of the entire T cell vector responding to the alloreactivity operator matrix of mHA-HLA complexes for a specific organ, including the negative values

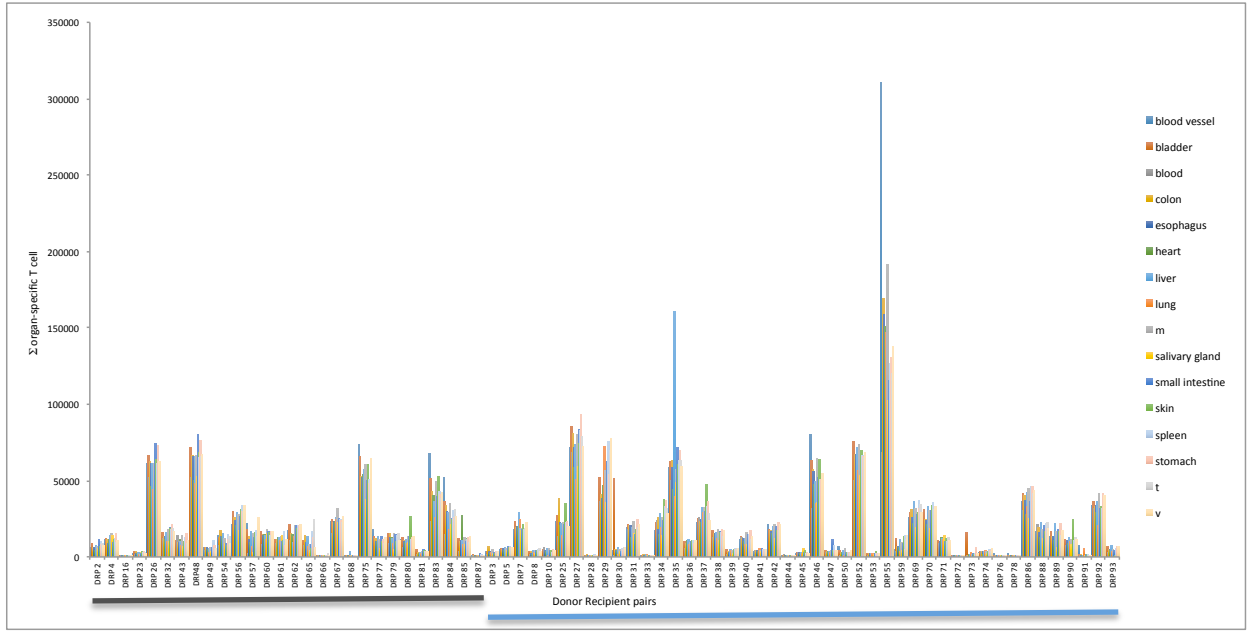

**Figure 3B & 3C.** Variation in the simulated alloreactive T cell counts observed for each organ examined between patients (3B) and within patients (3C). Note Log scale used in Figure 3B, but not in 3C (Y axis truncated at 130000).

**3B.**

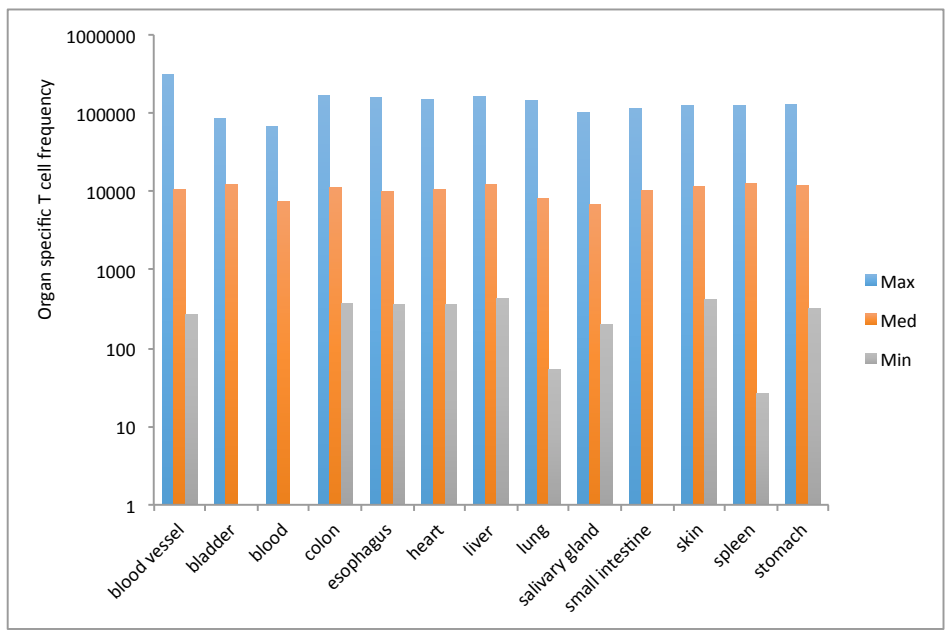

**3C.**

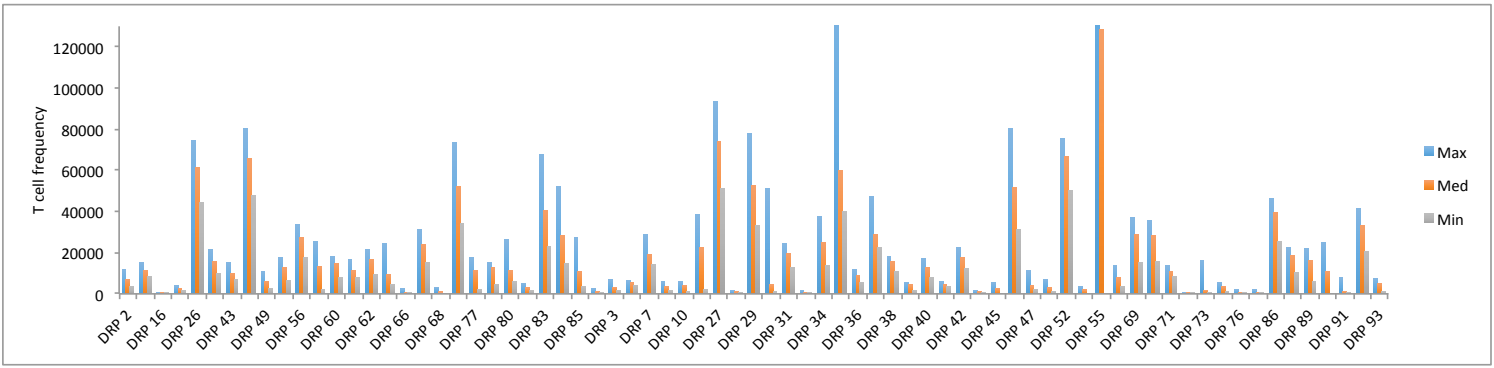

**Figure 3D.** Cumulative, simulated alloreactive T cell counts in recipients of MRD (Black line) and unrelated donors (Blue line)

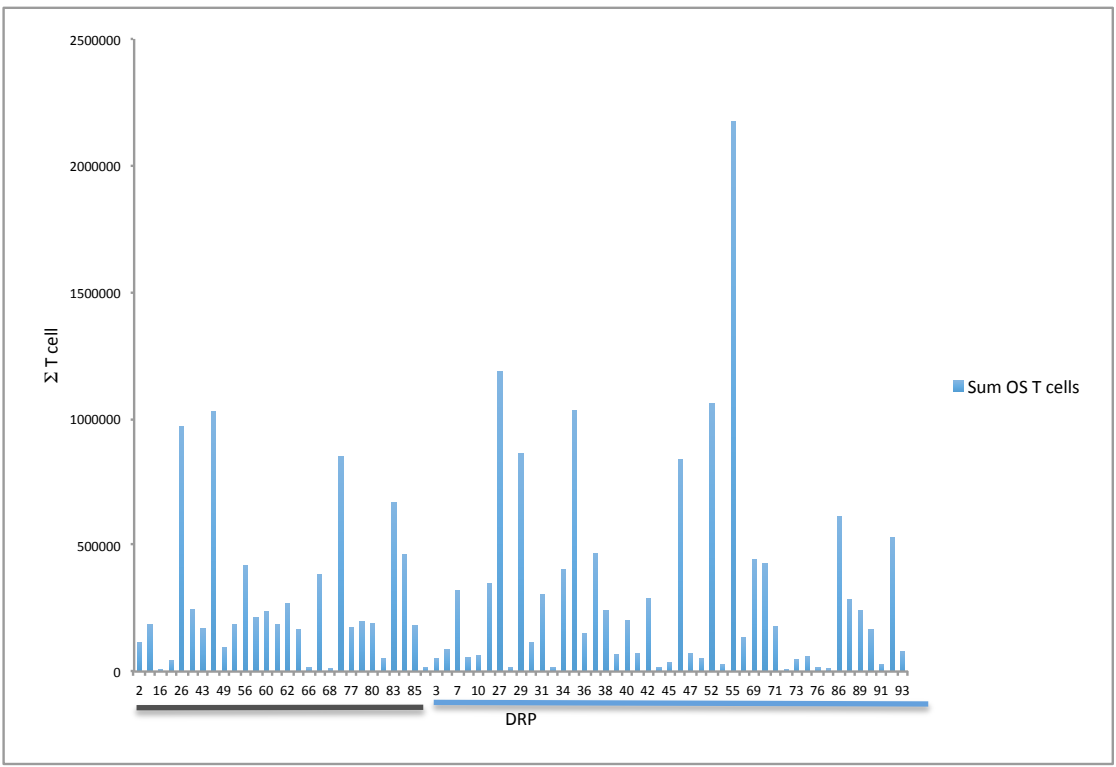

**Figure 4:** Alloreactivity model. Dot product of APC and T cell vectors recapitulates familiar antigen-challenge driven T cell proliferation response curve. A. Single T cell clone, B. Entire repertoire. C. Model illustrating the interaction between APC and T cells.

**4A**.

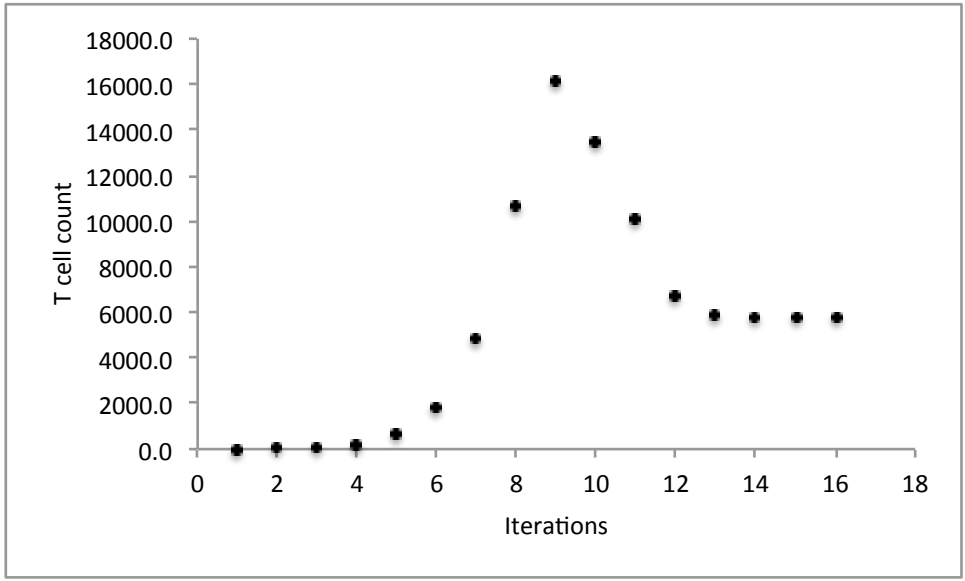

**4B**.

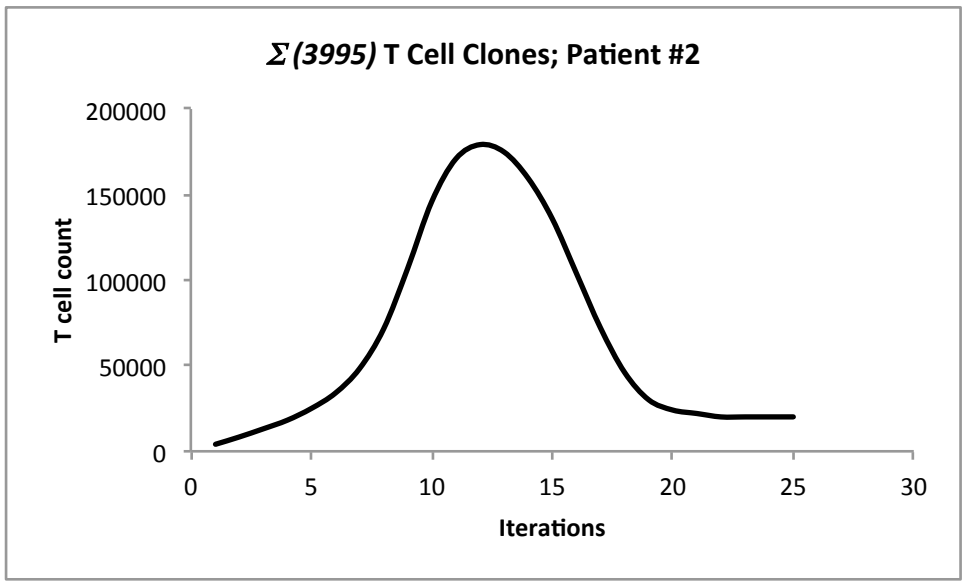

4C.

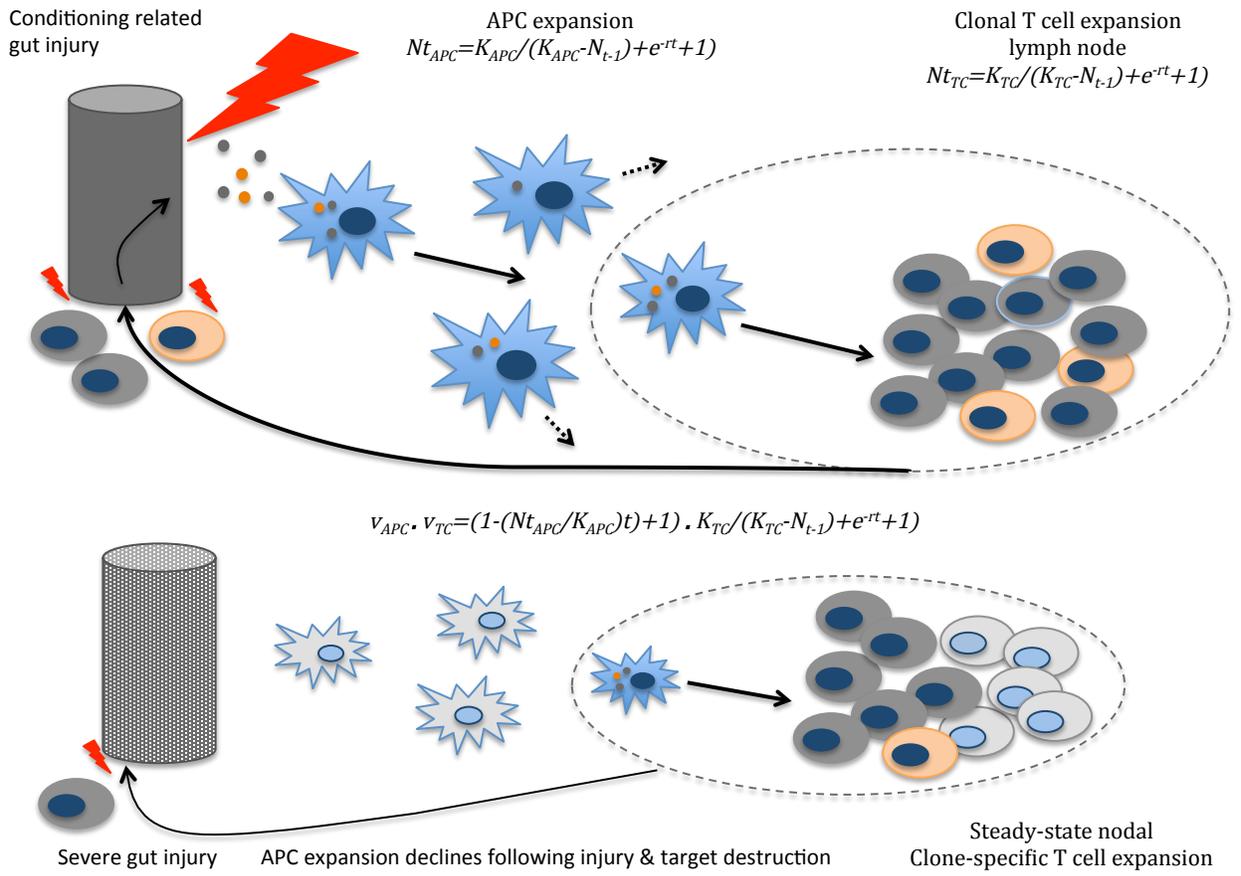

**Figure 5.** Modeling the effect of Treg on effector T cell growth: in the red curve *r* reduced at 21st iteration from 1 to 0.25; T cell population drops but then recovers slowly. In the blue curve *r* reduced at 25th iteration from 1 to 0.25 with direction reversal (from – to +), signifying anti-inflammatory effect supersedes pro-inflammatory effect.

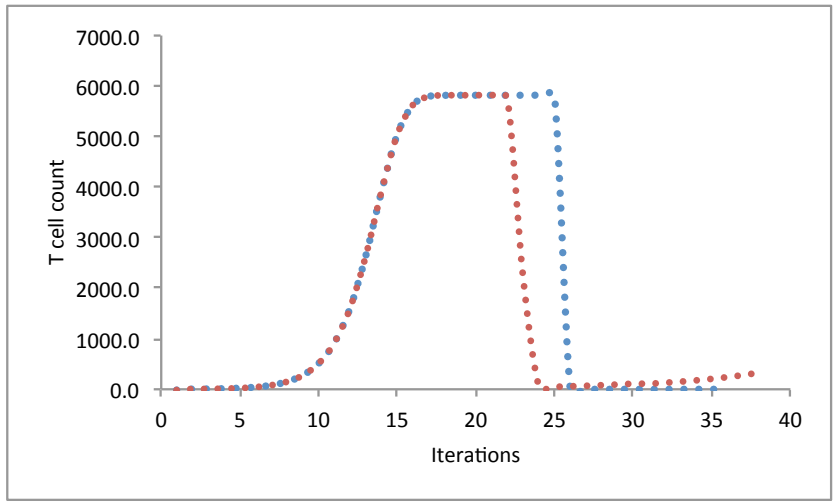

**Tables.**

**Table 1.** Exome sequencing and peptide results by donor type (n=77)

|  | Related[a] (n=27) | Unrelated (n=50) | *P*-value[b] |
|---|---|---|---|
| **Exome sequencing SNP differences** |  |  |  |
| Synonymous | 2,751 | 4,821 | <0.01 |
| Nonsynonymous | 2,490 | 4,287 | <0.01 |
| Conservative | 855 | 1,476 | <0.01 |
| Non-conservative | 1,635 | 2,811 | <0.01 |
| **HLA-Presented Peptides** |  |  |  |
| Peptides | 44,214 | 77,025 | <0.01 |
| Strong Binders | 838 | 1,160 | <0.01 |
| Presented Peptides | 3,626 | 5,386 | <0.01 |
| Percent of Strong Binders | 1.9 | 1.5 | 0.03 |
| Percent of Presented Peptides | 8.3 | 7.1 | 0.05 |
| Strong Binders/Presented Peptides | 22.6 | 21.1 | 0.08 |

[a]Excludes haploidentical patient

[b]Mean values compared using t-test for Equality of Means

**Supplementary Table 1.** Patient characteristics, excluding haploidental DRP

| Supplemental Table 1. | Patient Characteristics, n (%) | |
|---|---|---|
| Total Transplants | | 77 (100) |
| Gender | | |
|     Male/Female | | 43 (55.8)/34 (44.2) |
| Patient Age | | |
|     Median (range) | | 55.6 (21-73) |
| | | Total |
| **Age Distribution** | <40 | 12 (16) |
| | 40-59 | 43 (56) |
| | ≥60 | 22 (29) |
| **Donor** | MRD | 26 (34) |
| | MUD | 41 (53) |
| | MMRD | 1 (1) |
| | MMUD | 9 (12) |
| **Stem Cell Source** | Bone Marrow | 7 (9) |
| | Peripheral Blood | 70 (91) |
| **Diagnosis** | Acute lymphoid leukemia | 5 (6) |
| | Acute myeloid leukemia | 29 (38) |
| | Chronic lymphocytic leukemia and lymphomas | 15 (19) |
| | Multiple myeloma | 4 (5) |
| | Myelodysplastic syndromes | 24 (31) |
| **Conditioning Regimen** | Anti-thymocyte globulin (ATG) | 62 (81) |
| | Reduced Intensity | 46 (60) |
| |     ATG/TBI | 19 (25) |
| |     Busulfan/Fludarabine | 27 (35) |
| |     Fludarabine/Melphalan | 2 (77) |
| | Myeloablative | 31 (40) |
| |     Busulfan/Cyclophosphamide | 17 (22) |
| |     Cyclophosphamide/TBI | 11 (14) |
| |     Etoposide/TBI | 1 (1) |
| **GVHD prophylaxis** | Cyclosporin A/Methotrexate | 16 (21) |
| | Cyclosporin A/MMF | 4 (5) |
| | Tacrolimus/Methotrexate | 34 (44) |
| | Tacrolimus/MMF | 23 (30) |
| **GVHD, Overall** | Acute, Grades I-II | 20 (26) |
| | Acute, Grades III-IV | 14 (18) |
| | Chronic | 39 (51) |
| | Both, Acute + Chronic | 25 (32) |

**Supplementary Table 2:** Expected T cell counts per donor type by organ (presented in 1,000s of cells)

| Table 2 | | | | |
|---|---|---|---|---|
| | MRD (N=27) | | MUD (N=45) | |
| | Mean | SD | Mean | SD |
| Salivary Glands | 13.8 | 14.9 | 14.0 | 20.0 |
| Colon | 18.4 | 18.1 | 20.2 | 30.1 |
| Esophagus | 17.2 | 17.8 | 18.6 | 28.1 |
| Small Intestines | 19.1 | 20.3 | 19.9 | 25.7 |
| Stomach | 19.6 | 19.5 | 21.1 | 27.3 |
| Liver | 18.4 | 17.8 | 22.6 | 32.1 |
| Lung | 12.9 | 12.9 | 16.5 | 25.6 |
| Skin | 19.3 | 19.6 | 20.9 | 25.9 |
| Total | 138.8 | 139.3 | 154.8 | 210.5 |
| Variance Measures | | | | |
| Standard Deviation | 3.6 | 2.8 | 4.7 | 6.8 |
| Minimum | 22.7 | 21.0 | 27.3 | 37.2 |
| Maximum | 11.8 | 13.1 | 12.8 | 19.3 |
| Range | 10.9 | 8.7 | 14.4 | 21.1 |

**Supplementary Table 3.** Expected T cell counts by organ in patients with and without GVHD (presented in 1,000s of cells)

| Table 3 | | | | | | |
|---|---|---|---|---|---|---|
| | Overall (N = 73) | | No GVHD (N=26) | | GVHD (N=47) | |
| | Mean | SD | Mean | SD | Mean | SD |
| Salivary Glands | 13.7 | 18.1 | 10.3 | 11.9 | 15.6 | 20.6 |
| Colon | 19.3 | 26.0 | 15.3 | 15.1 | 21.4 | 30.3 |
| Esophagus | 17.9 | 24.5 | 13.6 | 14.4 | 20.2 | 28.5 |
| Small Intestines* | 19.3 | 23.6 | 14.4 | 14.9 | 22.1 | 27.0 |
| Stomach | 20.3 | 24.4 | 15.5 | 15.9 | 22.9 | 27.9 |
| Liver | 20.8 | 27.5 | 16.5 | 15.9 | 23.1 | 32.1 |
| Lung | 15.0 | 21.6 | 11.4 | 12.7 | 16.9 | 25.1 |
| Skin | 20.0 | 23.6 | 16.1 | 16.9 | 22.2 | 26.5 |
| Total* | 146.7 | 185.1 | 113.1 | 114.9 | 165.7 | 213.8 |
| *Indicates sample sizes of 72 for overall and 46 for GVHD group due to a missing value | | | | | | |

**Supplementary Table 4.** Estimated hazards ratios, 95% confidence intervals and p-values from Cox-proportional hazards models of t-cell counts and GVHD classification

| Table 4 | | | |
|---|---|---|---|
| | Hazard Ratio* | 95% CI | p-value |
| Salivary Glands | 1.01 | 0.99, 1.03 | 0.1184 |
| Colon | 1.01 | 0.99, 1.02 | 0.1479 |
| Esophagus | 1.01 | 0.99, 1.02 | 0.1039 |
| Small Intestines | 1.01 | 0.99, 1.02 | 0.0946 |
| Stomach | 1.01 | 0.99, 1.02 | 0.1055 |
| Liver | 1.01 | 0.99, 1.02 | 0.0546 |
| Lung | 1.01 | 0.99, 1.02 | 0.1248 |
| Skin | 1.01 | 0.99, 1.02 | 0.1442 |
| Total | 1.00 | 0.99, 1.01 | 0.0976 |
| **Standard Deviation** | **1.08** | **1.01, 1.15** | **0.0275** |
| Minimum | 1.01 | 0.99, 1.03 | 0.1049 |
| **Maximum** | **1.01** | **1.00, 1.02** | **0.0462** |
| **Range** | **1.02** | **1.00, 1.05** | **0.0276** |
| *Hazard ratio based on t-cell counts expressed in 1,000s | | | |

**Supplementary Table 5.** Estimated hazards ratios, 95% confidence intervals and p-values from Cox-proportional hazards models of t-cell counts and acute and chronic GVHD classification

| Table 5 | | | | | | |
|---|---|---|---|---|---|---|
| | Acute GVHD | | | Chronic GVHD | | |
| | Hazard Ratio* | 95% CI | p-value | Hazard Ratio* | 95% CI | p-value |
| Salivary Glands | 1.01 | 0.99, 1.03 | 0.3806 | 1.01 | 0.99, 1.03 | 0.1090 |
| Colon | 1.01 | 0.99, 1.02 | 0.4175 | 1.01 | 0.99, 1.02 | 0.2394 |
| Esophagus | 1.01 | 0.99, 1.02 | 0.3379 | 1.01 | 0.99, 1.02 | 0.1633 |
| Small Intestines | 1.01 | 0.99, 1.02 | 0.3880 | 1.01 | 0.99, 1.03 | 0.0788 |
| Stomach | 1.01 | 0.99, 1.02 | 0.4788 | 1.01 | 0.99, 1.02 | 0.1321 |
| Liver | 1.01 | 0.99, 1.02 | 0.1752 | **1.01** | **1.00, 1.03** | **0.0411** |
| Lung | 1.01 | 0.99, 1.02 | 0.2664 | 1.01 | 0.99, 1.03 | 0.2520 |
| Skin | 1.01 | 0.99, 1.02 | 0.3851 | 1.01 | 0.99, 1.02 | 0.1732 |
| Total | 1.00 | 0.99, 1.00 | 0.3290 | 1.00 | 1.00, 1.00 | 0.1202 |
| Standard Deviation | 1.07 | 0.99, 1.15 | 0.0618 | **1.08** | **1.01, 1.16** | **0.0404** |
| Minimum | 1.01 | 0.99, 1.03 | 0.3492 | 1.01 | 0.99, 1.03 | 0.1520 |
| Maximum | 1.01 | 0.99, 1.02 | 0.1526 | 1.01 | 0.99, 1.02 | 0.0586 |
| Range | 1.02 | 0.99, 1.05 | 0.0539 | **1.03** | **1.01, 1.05** | **0.0266** |
| *Hazard ratio based on t-cell counts expressed in 1,000s | | | | | | |

**Supplementary Figure 1:** Transition/transversion ratio for the WES data from all the patients.

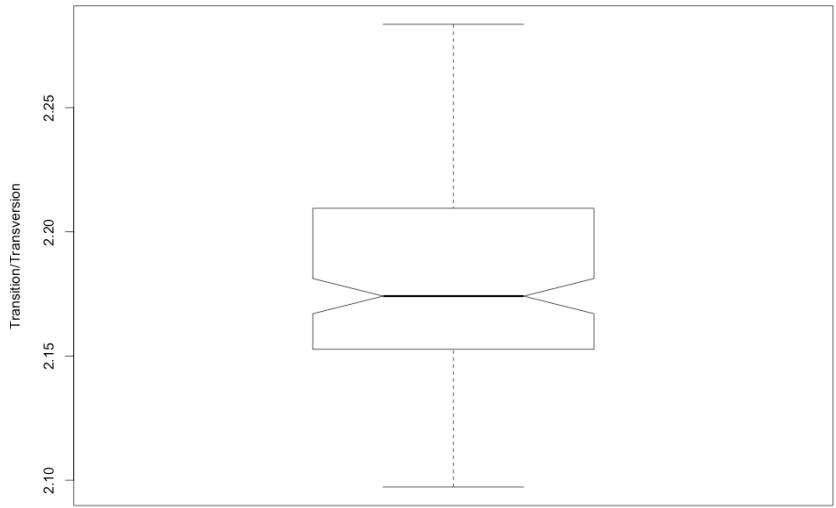

**Acknowledgements.** This work was supported by Massey Pilot Project Grant and an award from Virginia's Commonwealth Health Research Board (PI: MN). VK performed bioinformatic analysis of the sequencing data to identify unique peptides and their HLA binding affinity, as well as tissue expression. BA, SS wrote the program for performing calculations in MATLAB and performed vector-operator calculations presented in this paper. AS, DJK and TS collected and verified clinical outcome data. AS and RS performed statistical analysis. MS performed sequencing on samples identified and procured by CR. AS, CH and MJ-L created data files with unique peptides and HLA IC50 values. All the authors contributed to writing the manuscript.